\documentclass{bioinfo}
\copyrightyear{2006}
\pubyear{2006}

\begin{document}
\firstpage{1}

\title[Targeted Projection Pursuit]{Targeted Projection Pursuit for Gene Expression Data Classification and Visualisation}
\author[Faith \textit{et~al}]{Joe Faith\,$^{\rm a}$\footnote{to whom correspondence should be addressed}, Robert Mintram\,$^{\rm b}$, Maia Angelova\,$^{\rm a}$}
\address{$^{\rm a}$Northumbria University, Newcastle, UK, and $^{\rm b}$Bournemouth University, Bournemouth, UK}
\maketitle

\begin{abstract}
We present a novel method for finding low dimensional views of
high dimensional data: \emph{Targeted Projection Pursuit}. The
method proceeds by finding projections of the data that best
approximate a target view. Two versions of the method are
introduced; one version based on Procrustes analysis and one based
on a single layer perceptron. These versions are capable of
finding orthogonal or non-orthogonal projections respectively. The
method is quantitatively and qualitatively compared with other
dimension reduction techniques. It is shown to find
two-dimensional views that display the classification of cancers
from gene expression data with a visual separation equal to, or
better than, existing dimension reduction techniques.

\section{Contact:} \href{joe.faith@unn.ac.uk}{joe.faith@unn.ac.uk}
\end{abstract}

\section{Introduction}

This paper considers the problem of visualising classifications of
samples based on high dimensional gene expression data. There are
many powerful automatic techniques for analysing such data, but
visualisation represents an essential part of the analysis as it
facilitates the discovery of structures, features, patterns and
relationships, enables human exploration and communication of the
data and enhances the generation of hypotheses, diagnoses, and
decision making.

Visualising gene expression data requires representing the data in
two (or occasionally one or three) dimensions. Therefore
techniques are required to accurately and informatively show these
very high-dimensional data structures in low imensional
representations. In the particular case considered here, that of
showing the classification of gene expression data taken from
cancer samples, the most useful view will be one that clearly
shows the separation between classes, allowing the analyst to
easily identify outliers and cases of possible misdiagnosis, and
to visually compare particular samples.

There are many established techniques for viewing high-dimensional
data in lower dimensional spaces. Among these, multi-dimensional
scaling (MDS), including Sammon mapping, finds a view of the data
that best preserves the distances between points \citep{Ewi01};
VizStruct  is a technique based on radial coordinates
\citep{Zha04}; dendrograms may be used to linearly arrange and
display clustered gene expression data \citep{Eis98}; and
projection pursuit \citep{Lee05} finds linear projections that
optimise some measure of their quality (the 'projection pursuit
index').

Each of these techniques has limitations. MDS is a map-based,
rather than projection-based, technique in which adding  single
datum requires creating a new view of the entire set; thus it is
not possible to visualise the relationships of new or unclassified
samples to existing ones. VizStruct is not optimized for viewing
classifications of the data, and is also reliant on reducing the
dimensionality of the original data through some form of feature
selection. Dendrograms use linear arrangements of the data and so
are restricted to a single dimension for display. Projection
pursuit can optimize views for classification, and is based on a
linear projection, but is based on some form of search-based
optimisation and so also relies on feature selection to reduce the
size of the search space.

Here we present an alternative method for finding linear and
non-linear two-dimensional projections that yield views of the
data that are closest to a hypothesised optimal target. The method
is efficient and, in one version, not reliant on feature
selection. It is compared both quantitatively and subjectively
with existing techniques and is found to perform similarly to the
best of alternatives. When combined with other techniques it can
efficiently find views that are better than all alternatives and
which are close to the theoretical optimum.

\section{Targeted Projection Pursuit}\label{sec:target}

Conventional projection pursuit proceeds by searching the space of
all possible projections to find that which aximises an index that
measures the quality of each resulting view. In the case
considered here, a suitable index would measure the degree of
clustering within, and separation between, classes of points
\citep{Lee05}. Targeted projection pursuit, on the other hand,
proceeds by hypothesising an ideal view of the data, and then
finding a projection that best approximates that view. The
intuition motivating this technique is that the space of all
possible views of a high dimensional data set is extremely large,
so search-based methods of finding particular views may not be
effective. Hence the alternative technique is pursued of
suggesting an ideal view and then finding a nearest match.

Suppose $X$ is an $n \times p$ matrix that describes the
expression of $p$ genes in $n$ samples and $T$ is a $n \times 2$
matrix that describes a two-dimensional target view of those
samples. We require the $p \times 2$ projection matrix, $P$, that
minimises the size of the difference between the view resulting
from this projection of the data and our target:

\begin{equation}\label{eq:min}
    min \parallel T-XP \parallel
\end{equation}

where   $\parallel . \parallel$ denotes the Euclidean norm.

Two methods are considered for solving equation~(\ref{eq:min}),
depending on whether the projection matrix is required to be
orthogonal or not.

\subsection{Orthogonal Projections}\label{Orthogonal}

If we make the restriction that projection P is an
orthogonal-column matrix, then equation ~(\ref{eq:min}) is an
example of a Procrustes problem \citep{Gow04}, and a solution may
be found using the following version of the Singular Values
Decomposition (SVD) method presented by Golub and Loan
\citep{Gol96}(see \citet{Cox01} for a discussion of earlier
treatments).

Golub and Loan's method finds the $p \times p$ projection matrix,
$Q$, that best maps an $n \times p$ set of data, $X$, onto an
$n\times p$ target view, $S$, as follows:

\begin{equation}\label{eq:procrustes}
    Q = UV^{T}
\end{equation}

Where the superscript ${T}$ in equation~ (\ref {eq:procrustes})
denotes the transpose operator, and where $U$ and $V$ are the $p
\times p$ square matrices with orthogonal columns derived from the
SVD of $S^{T}X$:

\begin{equation}\label{eq:svd}
    S^{T}X  = UDV^{T}    \qquad \textrm{where $D$ is diagonal.}
\end{equation}

However if the target view, $T$, is $n \times 2$ then it can be
expanded to an $n \times p$ matrix, $S$ by padding with columns of
zeroes. And the required $p \times 2$ projection, $P$, can be
derived from $Q$ by taking just the first two columns.

Efficient methods for SVD are available in most common mathematical
and statistical packages such as MATLAB and R. Moreover the
complexity of calculating a SVD is dependent on the rank of the
matrix, {\it i.e} the number of linearly independent rows or
columns, rather than its absolute size. Thus, where the number of
samples is much less than the number of genes ($n\ll{p}$), then the
complexity of solving a Procrustes equation will end to be dependent
on the former rather than the latter. Hence this technique scales
extremely efficiently to high gene numbers.

\subsection{Non-Orthogonal Projections}\label{NonOrthogonal}

\begin{figure}
\centerline{\includegraphics[width=8cm]{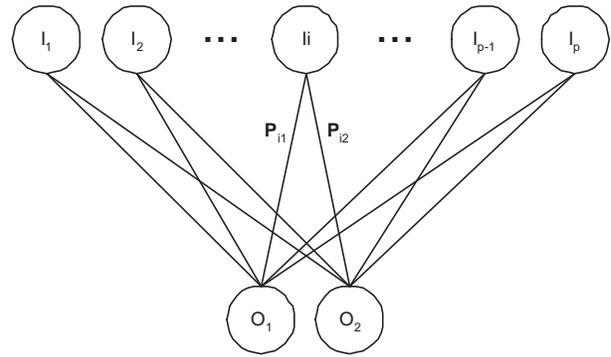}}
 \caption{Single Layer Perceptron for projecting $p$-dimensional data to $2$-dimensional view.}\label{fig:slp}
\end{figure}

If the projection $P$ is not required to be orthogonal then a
solution to equation~(\ref{eq:min}) may be found by training a
single layer perceptron with $p$ input units and 2 linear output
units (see Figure \ref{fig:slp}). Each of the $n$ data rows in $X$
are presented in turn, and standard back-propagation is used to
train the network to produce the corresponding row of $T$ in
response. Once converged, the network can be used to transform
data from the original gene-space to a 2 dimensional view, with
the weight of the connection from the $i^{th}$ input neuron to the
$j^{th}$ output neuron corresponding to the value of the
projection matrix $P_{ij}$.

\section{Targeted Projection Pursuit for Classification Visualisation}\label{sec:simplex}

Given a data set $X$ and a target view $T$, then the methods
described in Sections ~\ref{Orthogonal} and ~\ref{NonOrthogonal}
will find views of $X$ that approximate $T$. But what is the
appropriate target view when considering the classification of
gene expression data? If the samples are partitioned into $k$
known classes then the ideal view would be that in which the
classes are most clearly separated; that is where all members of
the same class are projected onto single points and where those
points are evenly spaced. Thus the ideal view is one in which all
the members of each class are projected onto a single vertex of a
geometric simplex.

\begin{figure}
\centerline{\includegraphics[width=8cm]{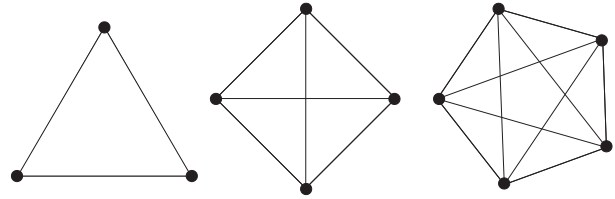}}
\caption{Graphs of the three-simpleces thru 5-
simplices.}\label{fig:simplex}
\end{figure}

The $k$-simplex, or hypertetrahedron, is the generalisation of an
equilateral triangle ($k=3$) or tetrahedron ($k=4$) to higher
dimensions. That is, the simplest possible polytope in any given
space, that in which all vertices are equidistant from each other.
The $k$-simplex itself is a polytope in $k-1$ dimensions, but
two-dimensional graphs of the three-simplices thru %? through
5-simplices
are shown in Figure \ref{fig:simplex}.

For example, given a set of samples taken from three classes over a large number of dimensions then
the ideal view of that data would approximate an equilateral triangle, with the samples of each class
clustered at the vertices, and hence would show the clustering within, and the separation between, classes.
Whether or not an accurate approximation to such a view can be found depends on how well separated the original data is.

The significance of using a $k$-simplex rather than just a regular polyhedron as our projection target can be
shown by considering the case of $k=4$. It may be supposed that the separation of classes could be effectively
shown by projecting the members of each class onto the vertices of a square.
However the vertices of a square are not equidistant: the two diagonal pairs of vertices are further apart
that the pairs of vertices on each edge. Therefore, using a square as a projection target would entail breaking
symmetry; effectively assuming that the pairs of classes that are mapped to the diagonally opposed vertices are
further separated than the other pairs. And this assumption may not be justified.
Mapping to the tetrahedron, on the other, makes no such assumption. Symmetry is not broken since each pair
of vertices are equally separated.

The procedure of mapping data onto a target view can also be considered in two other ways, other than the
geometric interpretation given above. First, as a set of binary classification problems; second as a spatial
classification problem.

First, note that the coordinates of the vertices of the k-simplex
can be generated by taking the rows of the $k$-dimensional identity
matrix; i.e. the unit diagonal matrix,

\begin{displaymath}
\left( \begin{array}{cccc}
1 & 0 & \ldots & 0 \\
0 & 1 & \ldots & 0 \\
\vdots & \vdots & \ddots & \vdots \\
0 & 0 & \ldots & 1 \\
\end{array} \right)
\end{displaymath}

Now, if $C_{i}$ is the set of members of the $i^{th}$ class (with complement $\bar{C_{i}}$),
then mapping the sample classes onto the $k$-simplex is equivalent to $k$ individual binary classification problems,
in which the $i^{th}$ column of our projection matrix, denoted as $P_{\cdot i}$, maps the members of $C_{i}$ to 1
and the members of $\bar{C_{i}}$  to 0. (A similar technique for reducing a multiclass classification problem to
multiple binary classifications is explored by \citealt{She06}.)

Alternatively, the projection onto a simplex can be thought of as mapping the original data
into `class space' -- a $k$ dimensional space in which the $j^{th}$  ordinate of the $i^{th}$  point represents
how closely the $i^{th}$  sample is correlated with the $j^{th}$ class.

\begin{figure}
\centerline{\includegraphics[width=8cm]{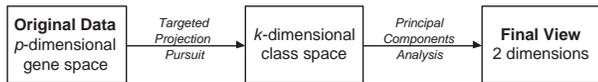}}
 \caption{Two-stage dimension reduction.}\label{fig:stage}
\end{figure}

Whether considered as a mapping onto a simplex, as a combination of
binary classification tasks, or as a mapping into class space, the
view of the data produced by targeted projection pursuit is
$k$-dimensional. Therefore where there are two or three classes the
result can be visualised directly. However, where $k>3$ then a
further dimension reduction step is required to view the data. Here
we use principal components analysis (PCA) on the rows of our
projection matrix, $P_{i\cdot}$, each a $k$-dimensional vector, to
find a lower dimensional projection that best preserves the
information in $P$. Thus we have a two-stage dimension reduction
process, each stage of which is based on a linear projection;
therefore the combined result is itself a linear projection (see
Figure \ref{fig:stage}).

\section{Method}
The targeted projection pursuit techniques outlined in Sections
\ref{sec:target} and \ref{sec:simplex} were tested for their ability
to produce two-dimensional views of data that clearly separate
sample classes. The techniques were tested on three publicly
available data sets, and the views were compared with the output
from standard dimension reduction techniques. The views of each data
set produced by each technique were tested both quantitatively and
qualitatively. The views were quantitatively compared in two ways:
first, by submitting them to a standard classification algorithm and
measuring the resulting generalisation performance; and, second, by
using a standard statistical measure of class separability. The
views were qualitatively compared by visual inspection.

The following dimension reduction techniques were compared:

\begin{itemize}
\item \textbf{SLP}: The result of targeted projection pursuit using a single layer linear perceptron network, followed by PCA.
\item \textbf{PRO}: The result of targeted orthogonal projection pursuit using the solution to a Procrustes equation, followed by PCA.
\item \textbf{PP}: The linear projection produced by search-based projection pursuit \citep{Lee05}
\item \textbf{SAM}: The result of a Sammon dimension-reduction non-linear mapping \citep{Ewi01}.
\item \textbf{VS}: The result of a VizStruct non-linear projection \citep{Zha04}.
\end{itemize}

All of these techniques, apart from PRO, required feature selection before processing, in which the top 50 most discriminatory genes were chosen on the basis of the ratio of their between-group to within-group sums of squares \citep{Dud02}.

The following data sets were used:

\begin{itemize}
\item \textbf{LEUK}: This dataset is the result of a study of gene expression in two types of acute leukemia: acute lymphoblastic leukemia (ALL) and acute myeloid leukemia (AML) \citep{Gol99}. The samples consist of 38 cases of B-cell ALL, 9 cases of T-cell ALL, and 25 cases of AML with the expression levels of 7219 genes measured. Note that, following Lee et al (2005), the B-cell and T-cell ALL samples are considered as separate classes.
\item \textbf{SRBCT}: This dataset comprises cDNA microarray analysis of small, round blue cell childhood tumors (SRBCT), including neuroblastoma (NB), rhabdomyosarcoma (RMS), Burkitt Lymphoma (BL; a subset of non-Hodgkin lymphoma) and members of Ewing' family of tumors (EWS). Expression levels from 6567 genes for 83 samples were taken \citep{Kha01}.
\item \textbf{NCI}: This dataset records the variation in gene expression among the 60 cell lines from the National Cancer Institute's anticancer drug screen \citep{Sch00}. It consists of 8 different tissue types where cancer was found: 9 breast, 5 central nervous system (CNS), 7 colon, 6 leukemia, 8 melanoma, 9 non-small-cell lung carcinoma (NSCLC), 6 ovarian, 2 prostate, 8 renal. 9703 cDNA sequences were used.
\end{itemize}

All data sets were normalised to zero mean and unit variance for each gene.

The classification algorithm used for the quantitative evaluation
was K Nearest Neighbours with $k=5$ ({\it i.e.} $5NN$). This choice
of algorithm was motivated by two considerations. The first is that
it is known to be effective at discriminating classes of tumour
using gene expression data \citep{Dud02}. The second consideration
is KNN is an instance- and distance-based measure in which the
classification of an instance is dependent on the classes of its
nearest neighbours. It is assumed that this measure would accord
better with human judgement than a probabilistic attribute-based
measure such as Na\"ive Bayes -- even though the latter may have
superior classification performance in some cases. The Weka
implementation of this algorithm was used \citep{Wit05}, tested
using 10-fold cross-validation, and a simple percentage accuracy
score found.

Note that the accuracy of classification using KNN for each view tested is not equivalent to a
true generalisation performance since the views were produced using the full data sets, rather than a training subset.
This is because it is the class separation within each view that is being tested, rather than the performance of the
classifier. Given a view, we would like to know how visually separated the classes in the data are -- operationalised
as classifier generalisation -- not which technique produces the best generalisation performance as a classifier.

The statistical measure of class separability used to compare views was the Linear Discriminant Analysis projection pursuit index ($I_{LDA}$) introduced by Lee \emph{et al} (2005), based on the ratio of between-groups to within-groups sums of squares. If $V_{ij}$ is the view of the $j^{th}$ member of the $i^{th}$ class then let

\begin{displaymath}
B = \sum_{i=1}^{k} n_i ( \bar{V_{i\cdot}} - \bar{V_{\cdot\cdot}} )( \bar{V_{i\cdot}} - \bar{V_{\cdot\cdot}} )^{T} \textrm{: between-group sum of squares}
\end{displaymath}
\begin{displaymath}
W = \sum_{i=1}^{k}\sum_{j=1}^{n_i} ( \bar{V_{ij}} - \bar{V_{i\cdot}} )( \bar{V_{ij}} - \bar{V_{i\cdot}} )^{T} \textrm{: within-group sum of squares}
\end{displaymath}

Thus $B$ is a measure of the variance of the centroids of the
classes, and $W$ is a measure of the variance of the instances
within each class. In order to get a projection pursuit index in
the range [0,1], with increasing values corresponding to
increasing class separation then, $I_{LDA}$, a version of Wilks
Lamda, a standard test statistic used in multivariate analysis of
variance, is used:

\begin{displaymath}
I_{LDA} = 1- \frac{|W|}{|W+B|}
\end{displaymath}

The R-code implementation of $I_{LDA}$ distributed by Lee \emph{et al} (2005) was used to measure the class seperability of the resulting views.

\section{Results}

\begin{figure}
\centerline{\includegraphics[width=8cm,height=6cm]{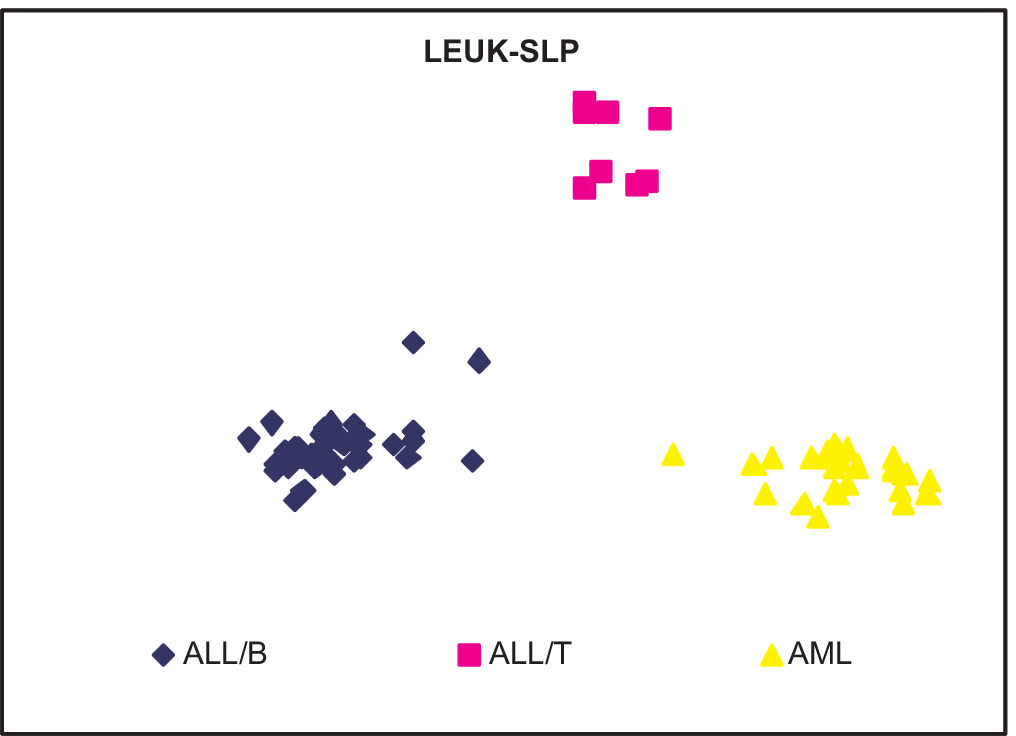}}
\caption{View of LEUK data set generated by SLP method.}\label{leuk-slp}
\end{figure}

\begin{figure}
\centerline{\includegraphics[width=8cm,height=6cm]{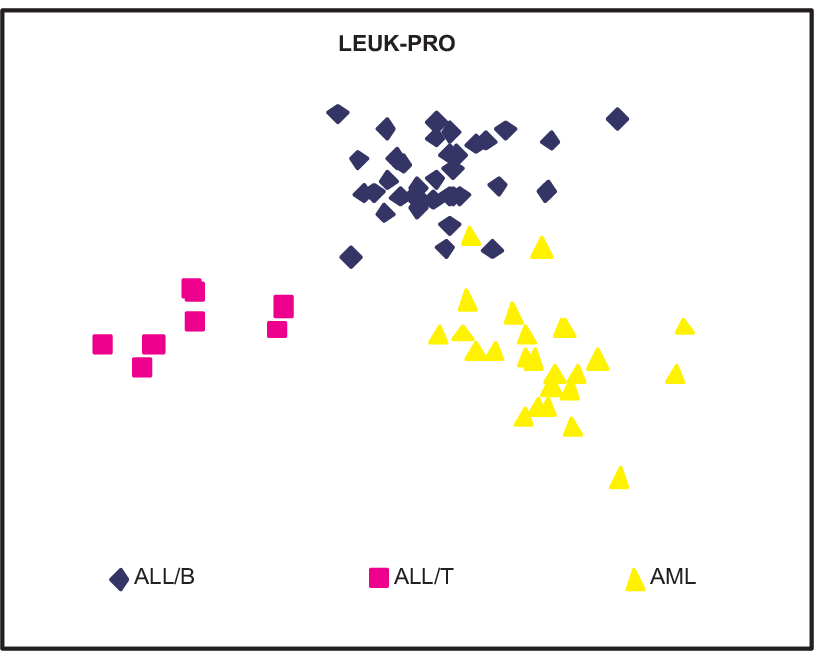}}
\caption{View of LEUK data set generated by PRO method.}\label{leuk-pro}
\end{figure}

\begin{figure}
\centerline{\includegraphics[width=8cm,height=6cm]{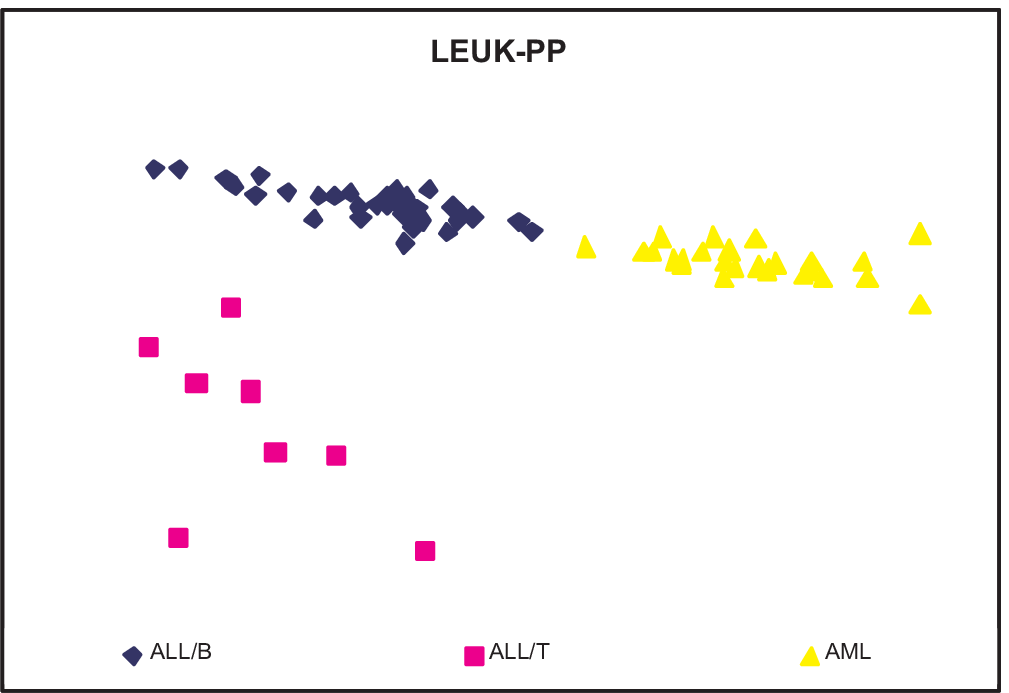}}
 \caption{View of LEUK data set generated by PP method.}\label{leuk-pp}
\end{figure}

\begin{figure}
\centerline{\includegraphics[width=8cm,height=6cm]{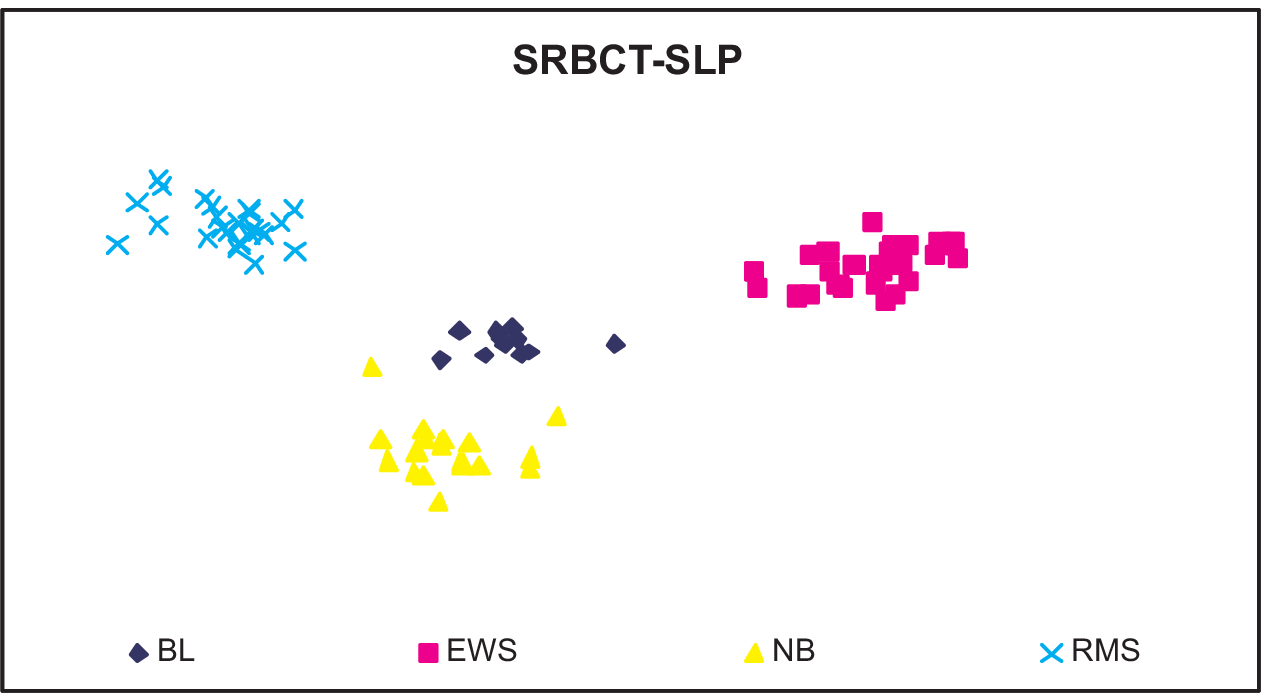}}
 \caption{View of SRBCT data set generated by SLP method.}\label{srbct-slp}
\end{figure}

\begin{figure}
\centerline{\includegraphics[width=8cm,height=6cm]{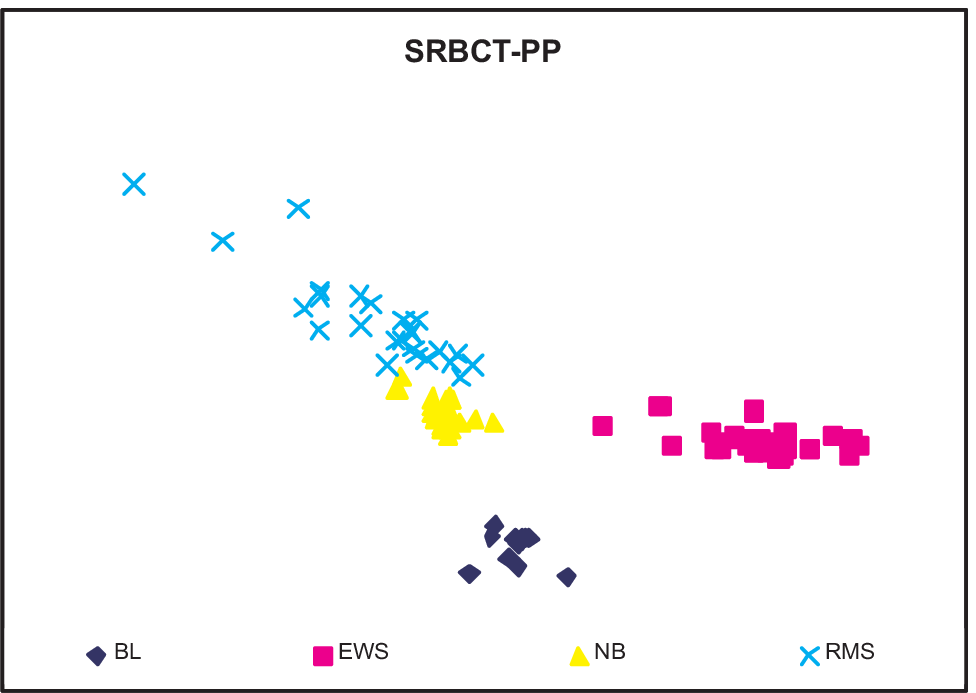}}
 \caption{View of SRBCT data set generated by PP method.}\label{srbct-pp}
\end{figure}

\begin{figure}
\centerline{\includegraphics[width=8cm,height=6cm]{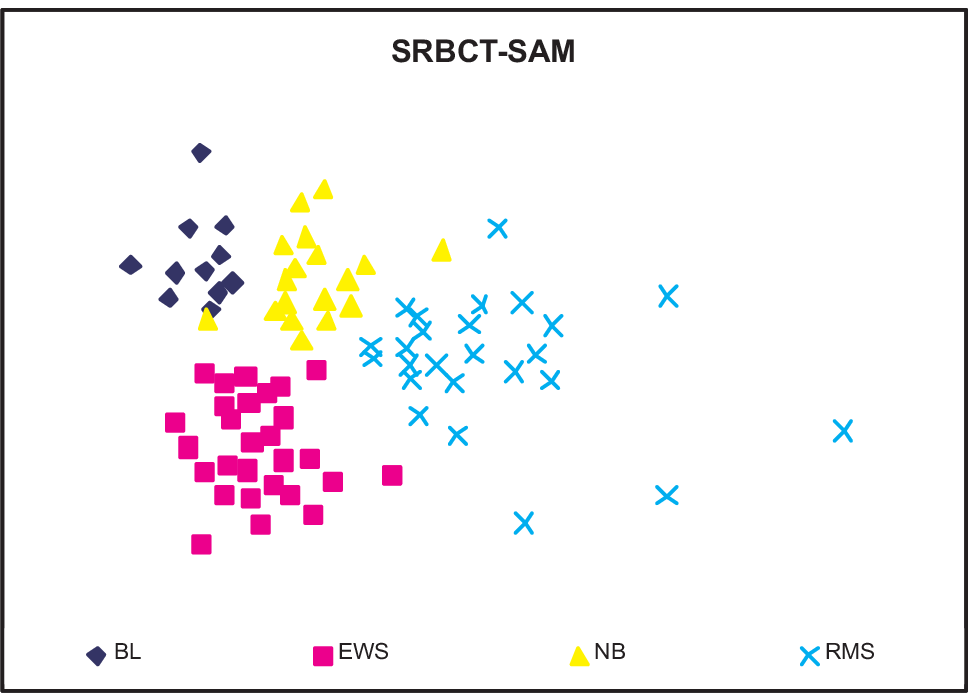}}
 \caption{View of SRBCT data set generated by SAM method.}\label{srbct-sam}
\end{figure}

\begin{figure}
\centerline{\includegraphics[width=8cm,height=6cm]{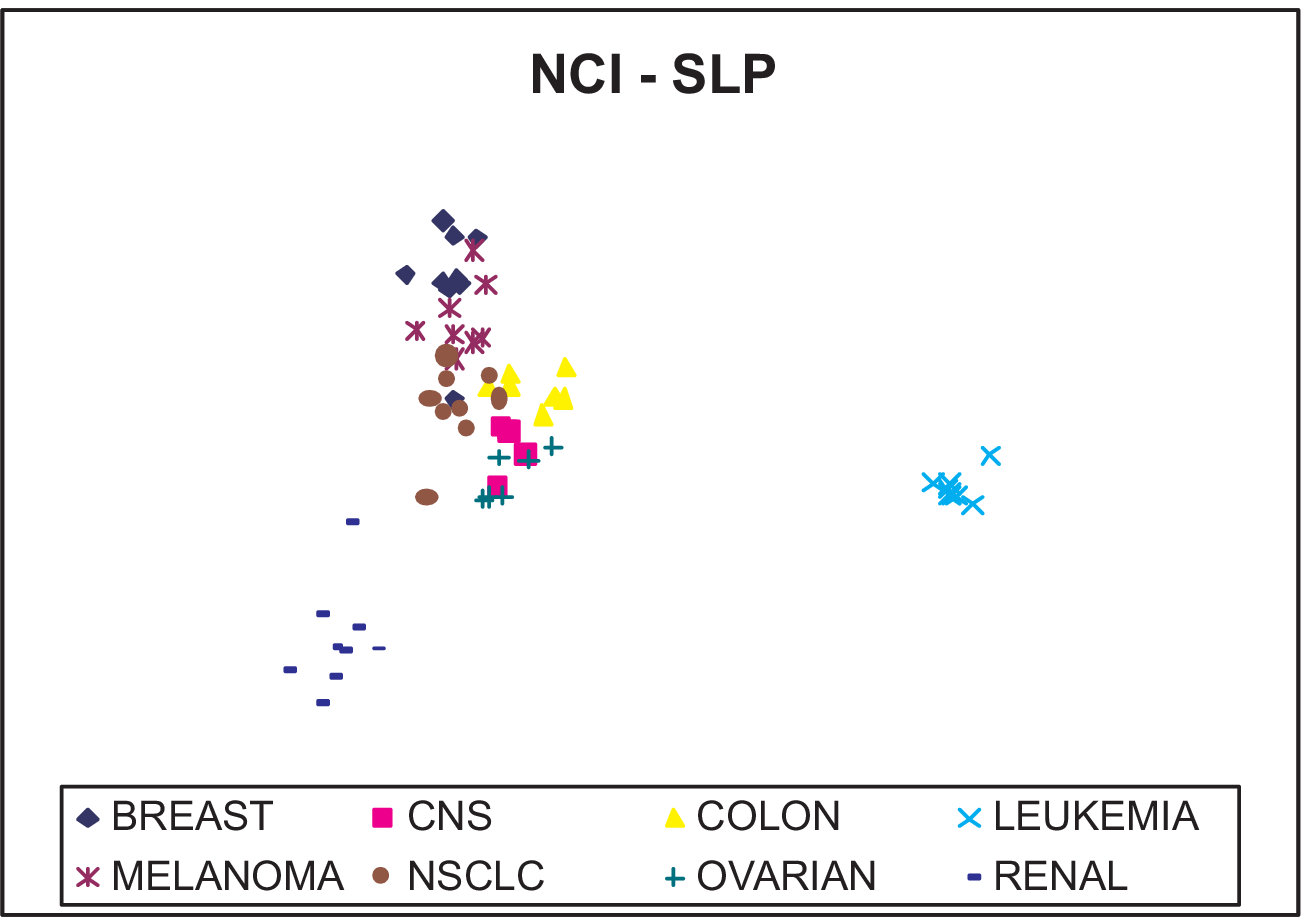}}
 \caption{View of NCI data set generated by SLP method.}\label{nci-slp}
\end{figure}

\begin{figure}
\centerline{\includegraphics[width=8cm,height=6cm]{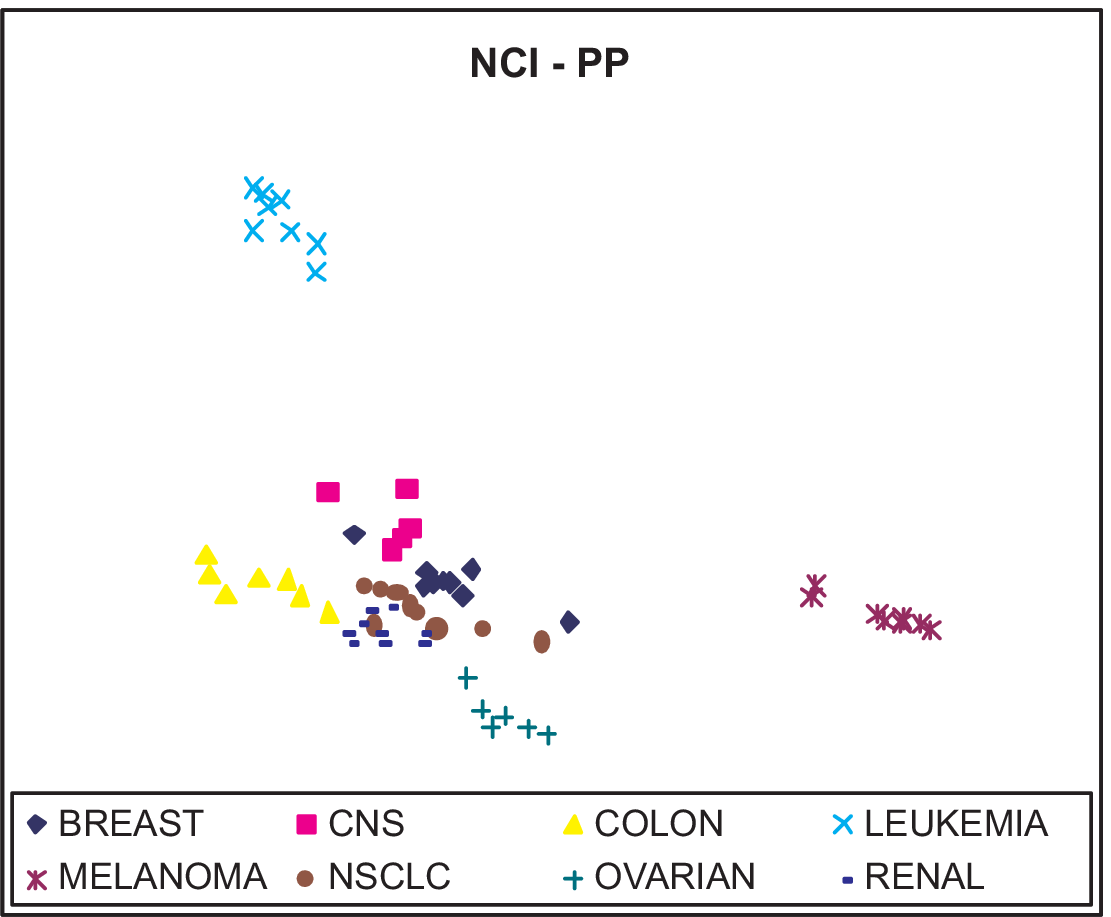}}
 \caption{View of NCI data set generated by PP method.}\label{nci-pp}
\end{figure}

\begin{figure}
\centerline{\includegraphics[width=8cm,height=6cm]{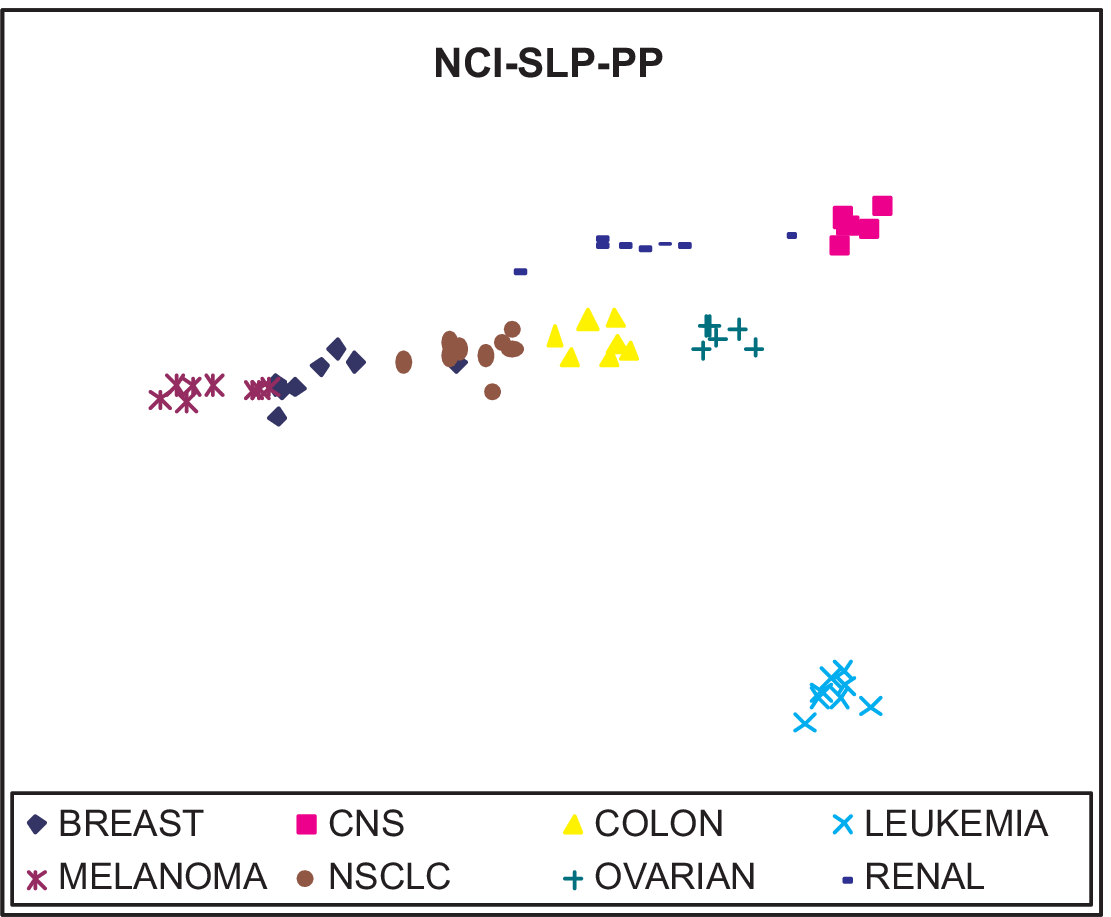}}
 \caption{View of NCI data set generated by hybrid SLP-PP method.}\label{nci-slp-pp}
\end{figure}

The quantitative comparison of the four projections on the three data sets is shown in Table \ref{quant}, and a sample of the resulting views are given in figures \ref{leuk-slp}-\ref{nci-slp-pp}.

\begin{table*}
\begin{center}
\caption{Comparison of class separability following dimension
reduction for visualisation. Each technique (SLP, PRO, PP, SAM and
VS) is evaluated on each data set (LEUK, SRBCT,NCI), and the
separability of the resulting view tested using both 5-Nearest
Neighbours classification ($5NN$, generalisation error in \%) and
a version of Wilks Lamda ($0<I_{LDA}<1$). The performance of the
hybrid technique (SLP-PP) on the NCI set is also shown (see
section \ref{sec:hybrid}).\label{quant}} {\begin{tabular}{l l|c
c|c c|c c}

\textbf{Data Set} & & \multicolumn{2}{c}{\textbf{LEUK}} & \multicolumn{2}{|c}{\textbf{SRBCT}} & \multicolumn{2}{|c}{\textbf{NCI}} \\
\hline
\textbf{Genes} & & \multicolumn{2}{c}{7129} & \multicolumn{2}{|c}{2308} & \multicolumn{2}{|c}{9712} \\
\textbf{Samples} & & \multicolumn{2}{c}{72} & \multicolumn{2}{|c}{83} & \multicolumn{2}{|c}{61} \\
\textbf{Classes} & & \multicolumn{2}{c}{3} & \multicolumn{2}{|c}{4} & \multicolumn{2}{|c}{8} \\
\hline
\multicolumn{2}{l|}{\textbf{Class Separation Measure}} & $I_{LDA}$ & $5NN$ & $I_{LDA}$ & $5NN$ & $I_{LDA}$ & $5NN$ \\
 & $SLP$ & .997 & 100 & .998 & 98.8 & .999 & 74.5 \\
 & $PRO$ & .945 & 94.4 & .971 & 96.4 & .967 & 54.1 \\
\textbf{Dimension Reduction Technique:} & $PP$ & .975 & 98.6 & .990 & 97.6 & .872 & 91.8 \\
 & $SAM$ & .959 & 97.2 & .911 & 95.2 & .958 & 67.2 \\
 & $VS$ & .951 & 95.8 & .637 & 56.6 & .818 & 42.6 \\
 & $SLP-PP$ & & & & & .999 & 88.5 \\
\hline

\end{tabular}}
\end{center}
\end{table*}

The first aspect of the results to note is that the choice of dimension-reduction technique can alter radically the resulting view of the data, judged both quantitatively and qualitatively. The structure and relationship between clusters appears very differently in each view, resulting in very different performances of classification algorithms. The choice of dimension reduction technique clearly matters in visualising high dimensional data such as gene expression data.

The second aspect to note is that quantitative measures such as $I_{LDA}$ or classification performance are not a reliable indicator of visual class separation. For example, the view of the NCI data set generation using SLP has an extremely high $I_{LDA}$ index of .999, but there is clear visual confusion between most of the classes (Figure \ref{nci-slp}). The discrepancy is probably due to the large distance between two of the classes (Leukemia and Renal) and the others; which will increase the measured variance between classes, but with little corresponding improvement in visual separation. As another example, SLP produced a view of the SRBCT data set with tightly clustered and clearly separated classes compared with PRO, SAM, and PP, and yet there is little difference in the generalisation performance of the $5NN$ classifier on these views (Figure \ref{srbct-slp}-Figure \ref{srbct-sam}).

Overall, VizStruct performed least well in separating classes. Although the difference between VizStruct and the other techniques was least for the low-$k$ case (LEUK), the difference became more marked as the number of classes increased. This poor performance is unsurprising, since this technique is not explicitly designed to accentuate classifications (though see Zhang {\it et al} (2004)).

The Sammon mapping performed well in separating classes, but its output was marked by the 'curse of dimensionality': in high dimensional spaces, the variance in distances between randomly distributed points decreases. Sammon mapping attempts to preserve the distance between data points, and hence the resulting views tend to be evenly distributed, with little 'bunching' of points belonging to a single class (Figure \ref{srbct-sam}). Classification algorithms may succeed in ascribing points to classes -- and hence the classification scores for SAM are similar to those for the linear mappings -- but this may not be an accurate reflection of the perceived class seperation.

For the low-$k$ data sets (LEUK and SRBCT), SLP-based targeted projection pursuit was the most effective method at separating classes, both quantitatively and visually, and the Procrustes method performs as well as conventional search-based projection pursuit. There are also qualitative similarities between the views produced by PRO and PP: for LEUK, the members of the ALL class are more similar to AML/B than either class is to AML/T; for SRBCT, NB are more closely related to RMS than either are to BL or EWS.

For the high-$k$, the conventional projection pursuit on a selected feature set outperformed either of the targeted methods; visually separating 4 classes compared to 2-3. In these cases there were also qualitative differences: PP most clearly distinguished leukemia, melanoma, and ovarian samples; Procrustes distinguished leukemia, breast, and colon samples; SLP distinguished leukemia and renal (Figure \ref{nci-slp}, Figure \ref{nci-pp}).

\subsection{Hybrid Projection Pursuit}
\label{sec:hybrid}

The targeted methods (SLP and PRO) performed relatively poorly in
the higher-$k$ case (NCI), compared with their success on the
lower-k cases (LEUK,SRBCT). This suggests that the drop in
performance is due to the second stage of the two-stage reduction
process, where PCA is used to reduce the dimensionality from
$k$-dimensional class space to the two-dimensional visualisation,
rather than the reduction from the original gene-space to
$k$-dimensional class-space (see Figure \ref{fig:stage}).

\begin{figure}
\centerline{\includegraphics[width=8cm]{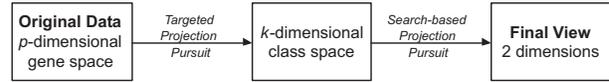}}
\caption{ Hybrid targeted and search-based projection pursuit.}\label{fig:hybrid}
\end{figure}

This hypothesis was tested by testing a hybrid dimension reduction
technique, in which SLP was used to reduce the dimensionality to k
and then search-based projection pursuit was used to find a
two-dimensional view of the result (Figure \ref{fig:hybrid}). Note
that the combined effect of this hybrid technique is still a
linear projection of the original data. This technique (SLP-PP)
was applied to the NCI data set and found to be highly effective
with a clear visual separation between five of the classes and
partial separation between remaining classes (Figure
\ref{nci-slp-pp}). Quantitatively, this hybrid had a better ILDA
measure than either technique used alone. It thus seems that a
limiting factor on search-based projection pursuit is the problem
of searching a very large space using a stochastic technique, such
as simulated annealing. Combining search-based projection pursuit
with SLP reduces the size of the space for the former task from
$50 \times 2$ dimensions to $8 \times 2$ in this case, and the
increase in performance is marked.

\section{Discussion}

The high dimensionality of microarray data introduces the need for visualisation techniques that can `translate' this data into lower dimensions without losing significant information, and hence assist with data interpretation. Many dimension reduction techniques are available, but in this paper we introduce the novel concept of targeted projection pursuit -- that is, finding views of data that most closely approximate a given target view -- and demonstrate the use of solutions of Procrustes equations and trained perceptron networks to achieve this end. In this particular case we explore the possibility of using targeted projection pursuit to find views that most clearly separate classified data sets.

Targeted projection pursuit was evaluated in comparison with three very different established dimension reduction techniques, on three publicly available data sets. The results are extremely encouraging. When discriminating a small number of cancer classes the performance of the technique matched or bettered that of established methods. When presented with a large number of classes (eight) the technique combined effectively with other existing techniques to produce near-optimal views of the data that could not be produced using any other means.

The technique is both simple to implement and powerful, scaling
well to large numbers of genes. The version of the technique
involving the targeted pursuit of orthogonal projections (PRO) is
able to handle an input dimensionality of tens of thousands of
genes without feature selection -- a capability unique amongst
those dimension reduction techniques tested. And informal trials
suggest that the neural network-based version is able to handle an
input dimensionality of many hundreds of genes without significant
loss in performance and using commodity hardware.

Note that the use of a target view does not constitute a limitation of the technique. The target plays the role of a hypothesis -- in this case that the samples can be classified based on gene expression levels -- and the resulting views illustrates how well the data meets that hypothesis. (And by using a fully-symmetrical simplex as the target view, no assumptions about the relationships between classes are made.) Other hypothesis-targets could be used in other cases, such as using a circular target to explore cyclical process in samples from a time-series, or a rectilinear target to explore the existence of simple linear relations. The same classification visualisation technique employed here to classify samples in gene-space could also be applied to the transpose problem; that of visualising the classification of genes on the basis of their expression profiles in varying conditions, and so explore relationships between gene function rather than between samples.

Targeted projection pursuit is a general purpose technique for finding views of data that approximate optimal targets. This paper discussed just one specific application to the problem of visualising classified microarray data. The authors are currently exploring other applications in visualising high dimensional biological data, including constructing a tool that would allow a user to explore the space of possible views of high dimensional data sets in intuitive and effective ways.

\section{Acknowledgement}
Thanks to Paul Vickers for comments on a previous version of this paper.

\end{document}